\documentclass[showpacs]{revtex4}
\usepackage{epsfig}
\def\beq{\begin{equation}}
\def\enq{\end{equation}}
\def\beqa{\begin{eqnarray}}
\def\enqa{\end{eqnarray}}

\def\MeV{\nobreak\,\mbox{MeV}}
\def\GeV{\nobreak\,\mbox{GeV}}

\def\qq{\lag\bar{q}q\rag}

\def\sss{\lag\bar{s}s\rag}
\def\qqs{\lag\bar{s}s\rag}

\def\Gd{\lag g^2G^2\rag}
\def\G3{\lag g^3G^3\rag}

\def\rh{\rho}
\def\si{\sigma}

\def\al{\alpha}
\def\be{\beta}
\def\alma{\alpha_{max}}
\def\almi{\alpha_{min}}
\def\bemi{\beta_{min}}
\def\mme{m_{D_s^*D_s^*}}

\def\lb{\label}
\def\nn{\nonumber}

\newcommand{\rag}{\rangle}
\newcommand{\lag}{\langle}

\begin{document}

\title{\sc A QCD sum rule calculation for the $Y(4140)$ narrow structure}
\author{Raphael M. Albuquerque}
\email{rma@if.usp.br}
\affiliation{Instituto de F\'{\i}sica, Universidade de S\~{a}o Paulo,
C.P. 66318, 05508-090 S\~{a}o Paulo, SP, Brazil}
\author{Mirian E. Bracco}
\email{bracco@uerj.br}
\affiliation{Faculdade de Tecnologia, Rodovia Presidente Dutra km
298, P\'olo Industrial,
CEP: 27.537-000, Resende, RJ, Brazil}
\author{Marina Nielsen}
\email{mnielsen@if.usp.br}
\affiliation{Instituto de F\'{\i}sica, Universidade de S\~{a}o Paulo,
C.P. 66318, 05389-970 S\~{a}o Paulo, SP, Brazil}

\begin{abstract}
We use the QCD  sum rules to evaluate the mass of a possible scalar mesonic 
state that couples to a molecular $D_{s}^{*}\bar{D}_s^{*}$ current. We find
a mass $\mme=(4.14\pm 0.09)$  GeV, which is in a excellent agreement
with the recently observed $Y(4140)$ charmonium state. 
We consider the contributions of condensates up to dimension eight, we
work at leading order in $\alpha_s$ and we keep terms which are linear in
the strange quark mass $m_s$. We also consider a molecular $D^{*}\bar{D}^{*}$ 
current and we obtain $m_{D^*{D}^*}=(4.13\pm 0.10)$, around 200 MeV above the 
mass of the $Y(3930)$ charmonium state. We conclude that it is possible to 
describe the $Y(4140)$ structure as a  $D_s^*\bar{D}_s^*$ molecular state.
\end{abstract}

\pacs{ 11.55.Hx, 12.38.Lg , 12.39.-x}
\maketitle


There is growing evidence that at least some of the new charmonium states 
recently discovery in the B-factories are non conventional $c\bar{c}$ states.
Some possible interpretations for these states are  mesonic molecules, 
tetraquarks,  or/and hybrid mesons. Some of these new mesons have their masses
very close to the meson-meson threshold like the $X(3872)$ \cite{belle1}
and the $Z^+(4430)$ \cite{belle2}. Therefore, a molecular interpretation
for these states seems natural. The most recent aquisiton for this list of
peculiar states is the narrow structure observed by the CDF Collaboration
in the decay $B^+\to Y(4140)K^+\to J/\psi\phi K^+$. The mass and width
of this structure is $M=(4143\pm2.9\pm1.2)~\MeV$, $\Gamma=(11.7^{+8.3}_{-5.0}
\pm3.7)~\MeV$ \cite{cdf}. Since the $Y(4140)$ decays into two 
$I^G(J^{PC})=0^-(1^{--})$ vector mesons, it has positive $C$ and $G$ parities.

There are already some theoretical interpretations for this structure. Its
interpretation as a conventional $c\bar{c}$ state is complicated because,
as pointed out by the CDF Collaboration \cite{cdf}, it lies well above the 
threshold for open charm decays and, therefore, a $c\bar{c}$ state with this 
mass would decay predominantly into an open charm pair with a large total 
width. In ref.~\cite{zhu}, the authors interpreted the $Y(4140)$ as the 
molecular partner of the charmonium-like state $Y(3930)$, which was observed 
by Belle and BaBar collaborations near the $J/\psi\omega$ threshold 
\cite{belba}. They concluded that the $Y(4140)$ is probably a $D_s^{*}
\bar{D}_s^{*}$ molecular state with $J^{PC}=0^{++}$ or $2^{++}$. 
In ref.~\cite{maha} they have 
interpreted the $Y(4140)$ as an exotic hybrid charmonium with $J^{PC}=1^{-+}$.

In this work, we use the QCD sum rules (QCDSR) \cite{svz,rry,SNB}, to
study the two-point function based on a  $D_s^*\bar{D}_s^*$ current with 
$J^{PC}=0^{++}$, to see if the new observed resonance structure, $Y(4140)$, 
can be interpreted as such molecular state.
In  previous calculations, the hidden charm mesons $X(3872),~Z^+(4430),~
Y(4260),~Y(4360),~Y(4660),~Z_1^+(4050)$ and $Z_2^+(4250)$ have been studied
using the QCDSR approach as tetraquark or molecular states 
\cite{x3872,molecule,lee,bracco,rapha,z12,zwid}.
In some cases a very good agreement with the experimental mass was obtained.


The starting point for constructing a QCD sum rule to evaluate the mass
of a hadronic state, $H$, is the correlator function
\beq
\Pi(q)=i\int d^4x ~e^{iq.x}\lag 0
|T[j_H(x)j_H^\dagger(0)]|0\rag,
\lb{2po}
\enq
where the current $j_H^\dagger$ creates the states with the quantum numbers
of the hadron $H$.
A possible current describing a $D_s^{*}\bar{D}_s^{*}$ molecular state with 
$I^GJ^{PC}=0^+0^{++}$ is
\beq
j=(\bar{s}_a\gamma_\mu c_a)(\bar{c}_b\gamma^\mu s_b)
\;,
\label{field}
\enq
where $a$ and $b$ are color indices.

The QCD sum rule is obtained by evaluating the correlation function in 
Eq.~(\ref{2po}) in two ways: in the OPE side, we
calculate the correlation function at the quark level in terms of
quark and gluon fields.  We work at leading order 
in $\alpha_s$ in the operators, we consider the contributions from 
condensates up to dimension eight and  we keep terms which are linear in
the strange quark mass $m_s$. In the  phenomenological side,
the correlation function is calculated by inserting intermediate states 
for the $D_s^{*}\bar{D}_s^{*}$ molecular scalar state.
Parametrizing the coupling of the scalar state,
$H=D_s^{*}\bar{D}_s^{*}$, to the current, $j$, in Eq.~(\ref{field}) in terms
of the parameter $\lambda$:
\beq\label{eq: decay}
\lag 0 |
j|H\rag =\lambda.
\label{lam}
\enq
the phenomenological side
of Eq.~(\ref{2po}) can be written as 
\beq
\Pi^{phen}(q^2)={\lambda^2\over
M_H^2-q^2}+\int_{0}^\infty ds\, {\rho^{cont}(s)\over s-q^2}, \lb{phe} 
\enq
where the second term in the RHS of Eq.(\ref{phe}) denotes higher scalar 
resonance contributions.

It is important to notice that  there is no one to one correspondence between
the current and the state, since the current
in Eq.~(\ref{field}) can be rewritten in terms of sum a over tetraquark type
currents, by the use of the Fierz transformation. However, the 
parameter $\lambda$, appearing in Eq.~(\ref{lam}), gives a measure of the 
strength of the coupling between the current and the state. 

The correlation function in the OPE side can be written as a
dispersion relation:
\beq
\Pi^{OPE}(q^2)=\int_{4m_c^2}^\infty ds {\rho^{OPE}(s)\over s-q^2}\;,
\lb{ope}
\enq
where $\rho^{OPE}(s)$ is given by the imaginary part of the
correlation function: $\pi \rho^{OPE}(s)=\mbox{Im}[\Pi^{OPE}(s)]$.

As usual in the QCD sum rules method, it is
assumed that the continuum contribution to the spectral density,
$\rho^{cont}(s)$ in Eq.~(\ref{phe}), vanishes bellow a certain continuum
threshold $s_0$. Above this threshold, it is given by
the result obtained with the OPE. Therefore, one uses the ansatz \cite{io1}
\beq
\rho^{cont}(s)=\rho^{OPE}(s)\Theta(s-s_0)\;,
\enq

To improve the matching between the two sides of the sum rule, we 
perfom a Borel transform. After transferring the continuum contribution to 
the OPE side, the sum rules for the scalar meson, considered as a scalar 
$D_s^*D_s^*$ molecule,
up to dimension-eight condensates, using factorization hypothesis, can
be written as:
\beq \lambda^2e^{-\mme^2/M^2}=\int_{4m_c^2}^{s_0}ds~
e^{-s/M^2}~\rho^{OPE}(s)\;, \lb{sr}
\label{sr1}
\enq
where
\beq
\rho^{OPE}(s)=\rho^{pert}(s)+\rh^{\sss}(s)
+\rh^{\lag G^2\rag}(s)+\rh^{mix}(s)+\rh^{\sss^2}(s)+\rh^{mix\sss}(s)\;,
\lb{rhoeq}
\enq
with
\beqa\label{eq:pert}
&&\rho^{pert}(s)={3\over 2^{9} \pi^6}\int\limits_{\almi}^{\alma}
{d\al\over\alpha^3}
\int\limits_{\bemi}^{1-\al}{d\be\over\be^2}(1-\al-\be)
\left[(\al+\be)m_c^2-\al\be s\right]^3\left(\left[(\al+\be)m_c^2-\al\be s
\over\beta\right]-4m_cm_s\right),
\nn\\
&&\rho^{\sss}(s)={3\sss\over 2^{5}\pi^4}\int\limits_{\almi}^{\alma}
{d\al\over\al}\left\{m_s{(m_c^2-\al(1-\al)s)^2\over1-\al}-
m_c\int\limits_{\bemi}^{1-\al}{d\be}\left[(\al+\be)m_c^2
-\al\be s\right]\times\right.\nn\\
&&\left.\times
\left[{(\al+\be)m_c^2-\al\be s\over\al\be}-{4m_sm_c\over\be}
\right]\right\},
\nn\\
&&\rho^{\lag G^2\rag}(s)={m_c^2\Gd\over2^{8}\pi^6}\int\limits_{\almi}^{
\alma}{d\al\over\alpha^3}\int\limits_{\bemi}^{1-\al}{d\be}(1-\al-\be)
\left[(\al+\be)m_c^2-\al\be s\right],
\nn\\
&&\rho^{mix}(s)=-{m_0^2\sss\over 2^{6}\pi^4}\left\{
3m_c\int\limits_{\almi}^{\alma}{d\al\over\al} [m_c^2-\al(1-\al)s]
-m_s(8m_c^2-s)\sqrt{1-4m_c^2/s}\right\}
,
\nn\\
&&\rho^{\sss^2}(s)={m_c\sss^2\over 8\pi^2}\left\{\sqrt{1-4m_c^2/s}\left(
2m_c-m_s\right)-m_sm_c^2\int_0^1{d\al\over\al}~\delta\left(
s-{m_c^2\over \al(1-\al)}\right)\right\},
\label{dim6}
\enqa
where the integration limits are given by $\almi=({1-\sqrt{1-
4m_c^2/s})/2}$, $\alma=({1+\sqrt{1-4m_c^2/s})/2}$, $\bemi={\al
m_c^2/( s\al-m_c^2)}$, and we have used $\lag\bar{s}g\si.Gs\rag=m_0^2\sss$. 
We have neglected the contribution of the dimension-six
condensate $\langle g^3 G^3\rangle$, since it is assumed to be suppressed
by the loop factor $1/16\pi^2$.  We also include a part
of the dimension-8 condensate contributions, related with the mixed 
condensate-quark condensate contribution:
\beqa
\rho^{mix\sss}(s)&=&-{m_cm_0^2\sss^2\over 16\pi^2}\int_0^1
d\al~\delta\left(s-{m_c^2\over \al(1-\al)}\right)\left[(2m_c-m_s)\left(1+{m_c^2
\over \al(1-\al)M^2}\right)
\right.
\nn\\
&-&\left.{5\over3}m_s\left(1-\al+{m_c^2\over\al M^2}+
{m_c^4\over2\al^2(1-\al)M^4}\right)\right].
\label{dim8}
\enqa
It is important to point out that a complete evaluation of the dimension-8 
condensate, and higher dimension condensates contributions, require more 
involved analysis \cite{bnp}, which is beyond the scope of 
this calculation.

To extract the mass $\mme$ we take the derivative of Eq.~(\ref{sr})
with respect to $1/M^2$, and divide the result by Eq.~(\ref{sr}).

For a consistent comparison with the results obtained for the other molecular
states using the QCDSR approach, we  have considered here the same values 
used for the quark masses and condensates  as in 
refs.~\cite{x3872,molecule,lee,bracco,rapha,z12,zwid,narpdg}:
$m_c(m_c)=(1.23\pm 0.05)\,\GeV $, $m_s=(0.13\pm 0.03)\,\GeV $,
$\lag\bar{q}q\rag=\,-(0.23\pm0.03)^3\,\GeV^3$, $\qqs=0.8\qq$,
$\lag\bar{s}g\si.Gs\rag=m_0^2\lag\bar{s}s\rag$ with $m_0^2=0.8\,\GeV^2$,
$\lag g^2G^2\rag=0.88~\GeV^4$.

\begin{figure}[h]
\centerline{\epsfig{figure=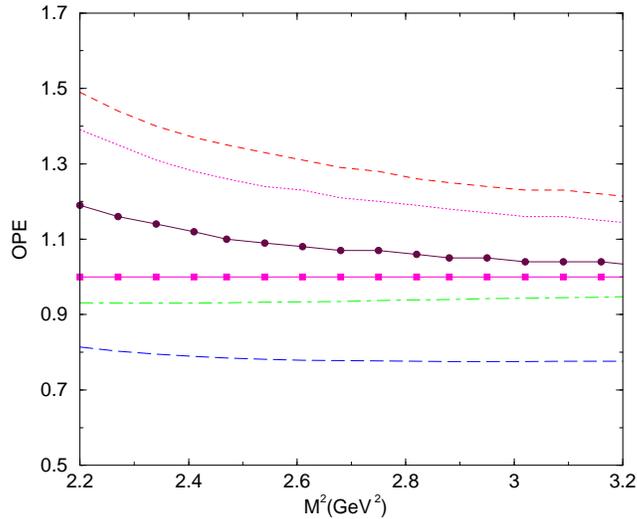,height=70mm}}
\caption{The OPE convergence for the $D_s^*D_s^*$ molecule in the region
$2.2 \leq M^2 \leq3.2~\GeV^2$ for $\sqrt{s_0} = 4.6$ GeV.  We plot the 
relative contributions starting with the perturbative contribution 
(long-dashed line), and each other line represents the relative contribution 
after adding of one extra condensate in the expansion: + $\sss$ (dashed line), 
+ $\langle g^2G^2\rangle$ (dotted line), + $m_0^2\sss$ (dot-dashed line), 
+ $\sss^2$ (line with circles), + $m_0^2\sss^2$ (line with squares).}
\label{figconv}
\end{figure}

The Borel window is determined by analysing the OPE convergence and the pole 
contribution. To determine the minimum value of the Borel mass we impose that
the contribution of the dimension-8 condensate should be smaller than 20\% of 
the total contribution.

In Fig.~\ref{figconv} we show the contribution of all the terms in the
OPE side of the sum rule. From this figure we see that for $M^2\geq 2.3$ 
GeV$^2$ the contribution of the dimension-8 condensate is less than 20\% of the
total contribution. Therefore, we  fix the lower
value of $M^2$ in the sum rule window as $M^2_{min}= 2.3$ GeV$^2$.

The maximum value of the Borel mass is 
determined by imposing that the pole contribution must be bigger than the 
continuum contribution. In Table I we show the values of $M^2_{max}$.
In our numerical analysis, we will consider the range of $M^2$ values
from 2.3 $\GeV^2$ until the one allowed by the pole dominance criterion given
in Table I.


\begin{center}
\small{{\bf Table I:} Upper limits in the Borel window for the $D_s^*D_s^*$ 
state obtained from the sum rule for different values of $\sqrt{s_0}$.}
\\
\begin{tabular}{|c|c|}  \hline
$\sqrt{s_0}~(\GeV)$ & $M^2_{max}(\GeV^2)$  \\
\hline
 4.4 & 2.49 \\
\hline
 4.5 & 2.68 \\
\hline
4.6 & 2.87 \\
\hline
4.7 & 3.06 \\
\hline
\end{tabular}\end{center}

The continuum threshold is a parameter of the 
calculation which, in general, is connected to the mass of the studied state,
$H$, by the relation $s_0\sim(m_H+0.5~\GeV)^2$. Therefore, to choose
a good  range to the value of $s_0$ we extract the mass from the sum rule, 
for a given $s_0$, and accept such value if the obtained mass is in the range 
0.4 GeV to 0.6 GeV smaller than $\sqrt{s_0}$. Using these criteria,
we obtain $s_0$ in the range $4.5\leq \sqrt{s_0} \leq4.7$ GeV.
However, because of the complex spectrum of the exotic states, some times 
lower continuum threshold values are favorable in order to completely 
eliminete the continuum above the resonance state. Therefore,  here
we will also include the result for the $D_s^*D_s^*$ meson mass for 
$\sqrt{s_0}=4.4\GeV$.

\begin{figure}[h]
\centerline{\epsfig{figure=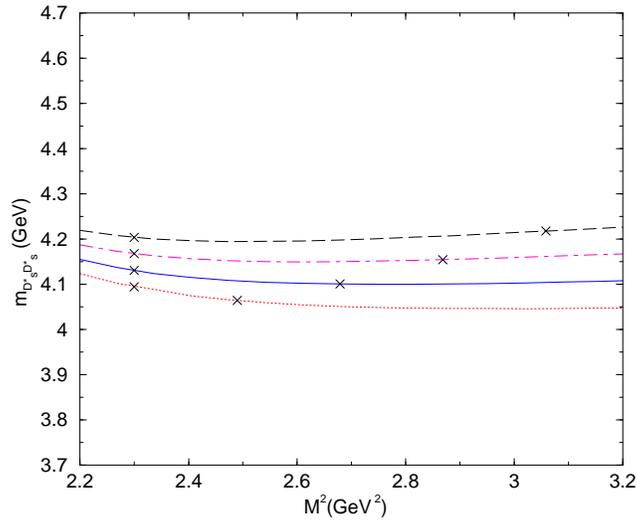,height=70mm}}
\caption{The $D_s^*D_s^*$ meson mass as a function of the sum rule parameter
($M^2$) for $\sqrt{s_0} =4.4$ GeV (dotted line), $\sqrt{s_0} =4.5$ GeV 
(solid line), $\sqrt{s_0} =4.6$ GeV 
(dot-dashed line) and $\sqrt{s_0} =4.7$ GeV (dashed line). The crosses
indicate the upper and lower limits in the Borel region.}
\label{figmz1}
\end{figure}

In Fig.~\ref{figmz1}, we show the $D_s^*D_s^*$ meson mass, for different values
of $\sqrt{s_0}$, in the relevant sum rule window, with the upper and
lower validity limits indicated.  From this figure we see that
the results are very stable as a function of $M^2$.
We see also that for $\sqrt{s_0}=4.4$ GeV, we get a very narrow Borel window, 
and for $\sqrt{s_0}=4.3$ GeV there is no allowed Borel window.

Using the Borel window, for each value of $s_0$, to evaluate the mass of the
$D_s^*D_s^*$ meson and then varying the value of the continuum threshold in
 the range $4.4\leq \sqrt{s_0} \leq4.7$
GeV, we get $\mme = (4.14\pm0.08)~\GeV$.

Up to now we have kept the values of the quark masses and condensates fixed.
To check the dependence of our results with these values we fix 
$\sqrt{s_0}=4.55~\GeV$ and vary the other parameters
in the ranges: $m_c=(1.23\pm0.05)~\GeV$, $m_s=(0.13\pm 0.03)\,\GeV $,
$\lag\bar{q}q\rag=\,-(0.23\pm0.03)^3\,\GeV^3$, $m_0^2=(0.8\pm0.1)\GeV^2$. 
In our calculation we have assumed the factorization hypothesis. However, it 
is important to check how a violation of the factorization hypothesis would
modify our results. For this reason we multiply $\sss^2$ in Eqs.~(\ref{dim6}) 
and (\ref{dim8}) by a factor $K$ and we vary $K$ in the range 
$0.5\leq K\leq2$.

\begin{center}
\small{{\bf Table II:} Values obtained for $\mme$, in the Borel window 
$2.38\leq M^2\leq 2.72~\GeV^2$, when the parameters vary in the ranges
showed.}
\\
\begin{tabular}{|c|c|}  \hline
parameter & $\mme~(\GeV)$  \\
\hline
$m_c=(1.23\pm0.05)~\GeV$  & $4.15\pm0.08$ \\
\hline
$m_s=(0.13\pm 0.03)\,\GeV $ &$4.14\pm0.02$  \\
\hline
$\lag\bar{q}q\rag=\,-(0.23\pm0.03)^3\,\GeV^3$ & $4.14\pm0.03$ \\
\hline
$m_0^2=(0.8\pm0.1)\GeV^2$ & $4.15\pm0.07$ \\
\hline
$0.5\leq K\leq2$ & $4.14\pm0.03$ \\
\hline
\end{tabular}\end{center}

Taking into account the incertainties given above we finally arrive at
\beq
\mme = (4.14\pm0.09)~\GeV,
\label{ymass}
\enq
in an excellent agreement with the mass of the narrow structure $Y(4140)$
observed by CDF.

One can also deduce, from Eq.~(\ref{sr1}), the parameter 
$\lambda$ defined in Eq.~(\ref{lam}). We get:
\beq
\lambda = \left(4.22\pm0.83
\right)\times 10^{-2}~\GeV^5,
\label{la1}
\enq

\begin{figure}[h]
\centerline{\epsfig{figure=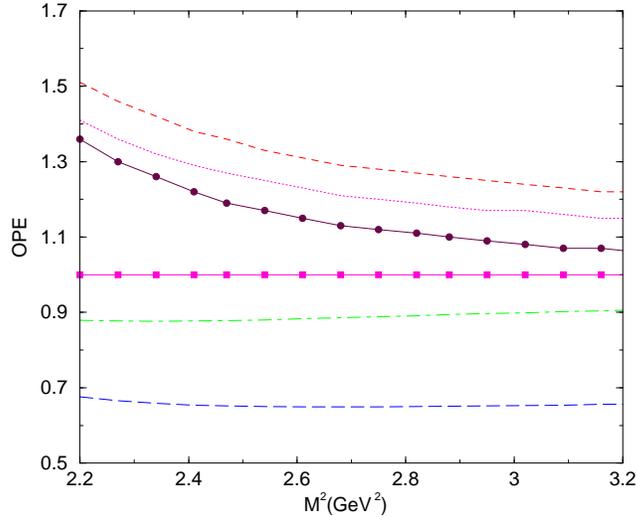,height=70mm}}
\caption{Same as Fig. 1 for the $D^*D^*$ state for $\sqrt{s_0}=4.6~\GeV$.}
\label{opedd}
\end{figure}

From the above study it is very easy to get results for the $D^*\bar{D}^*$ 
molecular state with $J^{PC}=0^{++}$. For this we only have to take
$m_s=0$ and $\sss=\qq$ in Eqs.~(\ref{dim6}), (\ref{dim8}). This study was 
already done in ref.~\cite{z12} considering $4.5\leq\sqrt{s_0}\leq
4.7~\GeV$. Although in the case of the $D^*D^*$ scalar molecule we get a 
worse Borel convergence than for the $D_s^*D_s^*$ scalar molecule, as can be 
seen by  Fig.~\ref{opedd}, there is still a good OPE convergence for 
$M^2\geq 2.5~\GeV^2$. 

If we allow also for the $D^*\bar{D}^*$ molecule values of the
continuum threshold in the range $4.4\leq \sqrt{s_0}\leq4.7~\GeV$
we get $m_{D^*D^*}=(4.13\pm0.11)~\GeV$. Therefore, 
from a QCD sum rule study, the difference between the masses of the states
that couple with scalar $D_s^*\bar{D}_s^*$ and  $D^*\bar{D}^*$ currents, is 
consistent with zero. The mass obtained with the $D^*\bar{D}^*$ scalar
current is about 100 MeV above the $D^*D^*(4020)$ threshold. This could be
an indication that there is a  repulsive interaction between the two $D^*$ 
mesons. Strong interactions effects might lead to repulsive interactions that
could result in a virtual state above the threshold. Therefore,
this structure may or may not indicate a resonance. However, considering the 
errors, it is not compatible with the observed $Y(3940)$ 
charmonium-like state.

\begin{figure}[h]
\centerline{\epsfig{figure=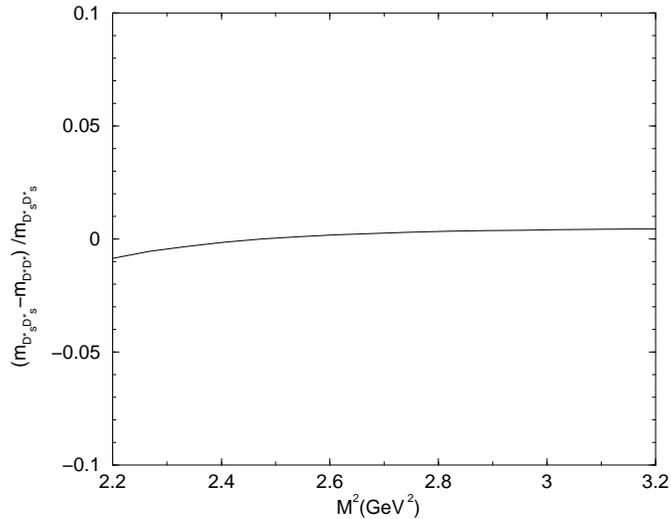,height=70mm}}
\caption{The relative ratio between the masses of the scalar molecular states
$\mme$ and $m_{D^*D^*}$ for $\sqrt{s_0}=4.55~\GeV$.}
\label{dif}
\end{figure}

In Fig.~\ref{dif} we show the relative ratio $(\mme-m_{D^*D^*})/\mme$ 
as a function of the Borel mass for $\sqrt{s_0}=4.55~\GeV$. From this figure
we can see that the ratio is very stable as 
a function of $M^2$ and the difference between the masses is smaller than 
0.5\%. Although the ratio is shown for $\sqrt{s_0}=4.55~\GeV$, the result
is indiscernible from the one shown in  Fig.~\ref{dif} for other values 
of the continuum threshold in the range 
$4.4\leq\sqrt{s_0}\leq4.7~\GeV$.

This result for the mass difference is completely unexpected since, in 
general, each strange quark adds approximately 100 MeV to the mass of the 
particle. Therefore, one would naively expect that the mass of the 
$D_s^*D_s^*$ state should be around 200 MeV heavier than the mass of the
$D^*D^*$ state. This was, for instance, the result obtained in 
ref.~\cite{rapha} for the vector molecular states $D_{s0}\bar{D}_s^*$ and 
$D_{0}\bar{D}^*$, where the masses obtained were:
$m_{D_{s0}\bar{D}_s^*}=  (4.42\pm0.10)~\GeV$ and
$m_{D_{0}\bar{D}^*}=  (4.27\pm0.10)~\GeV~$.

For the value of the parameter $\lambda$ we get:
\beq
\lambda_{D^*D^*} = (4.20\pm0.96)\times 10^{-2}~\GeV^5.
\label{la2}
\enq
Therefore, comparing the results in Eqs.~(\ref{la1}) and (\ref{la2})
we conclude that the currents couple with similar strength to the 
corresponding states, and that both, $D_s^*\bar{D}_s^*$ and $D^*\bar{D}^*$ 
scalar molecular states have masses compatible with the recently observed
$Y(4140)$ narrow structure. However, since the $Y(4140)$ was observed in the 
decay $Y(4140)\to J/\psi\phi$,  the  $D_s^*\bar{D}_s^*$ assignment
is more compatible with its quark content.

In conclusion, we have presented a QCDSR analysis of the two-point
function for possible  $D_s^*\bar{D}_s^*$ and $D^*\bar{D}^*$ molecular 
states 
with $J^{PC}=0^{++}$. Our findings indicate that the $Y(4140)$ narrow 
structure observed by  the CDF Collaboration
in the decay $B^+\to Y(4140)K^+\to J/\psi\phi K^+$ can be very well described
by using a scalar $D_s^*\bar{D}_s^*$ current.
Although the authors of ref.~\cite{zhu} interpreted the $Y(4140)$ as 
a $D_s^{*}\bar{D}_s^{*}$ molecular scalar state and the $Y(3930)$ as 
a $D^{*}\bar{D}^{*}$ molecular scalar state, we have obtained similar masses
for the states that couple with the scalars $D_s^{*}\bar{D}_s^{*}$ and 
$D^{*}\bar{D}^{*}$ currents. Therefore, from a QCD sum rule point of view, the
charmonium-like state $Y(3930)$, observed by Belle and BaBar Collaborations,
has a mass around 200 MeV smaller than the state that couples with a
$D^{*}\bar{D}^{*}$ scalar current and, therefore, can not be well described
by such a current.

While this work has been finalized, a similar calculation was presented in 
ref.~\cite{wang}. However, the author of ref.~\cite{wang} arrived to a 
different conclusion.

\section*{Acknowledgements}

We would like to thank F.S. Navarra for fruitiful conversations.
 This work has been partly supported by FAPESP and CNPq-Brazil.


\end{document}